# Theoretical analysis of doped graphene as cathode catalyst in Li-O$_2$ and Na-O$_2$ batteries – the impact of the computational scheme

Katarina A. Novčić[1], Ana S. Dobrota[1], Milena Petković[1], Börje Johansson[2,3,4], Natalia V. Skorodumova[2,3], Slavko V. Mentus[1,5], Igor A. Pašti[1,2,*]

[1]*University of Belgrade − Faculty of Physical Chemistry, Belgrade, Serbia*
[2]*Department of Materials Science and Engineering, School of Industrial Engineering and Management, KTH-Royal Institute of Technology, Stockholm, Sweden*
[3]*Department of Physics and Astronomy, Uppsala University, Uppsala, Sweden*
[4]*Humboldt University, Physics Department, Berlin, Germany*
[5]*Serbian Academy of Sciences and Arts, Belgrade, Serbia*

**Abstract**

Understanding the reactions in M-O$_2$ cells (M = Li or Na) is of great importance for further advancement of this promising technology. Computational modelling can be helpful along this way, but an adequate approach is needed to model such complex systems. We propose a new scheme for modelling processes in M-O$_2$ cells, where reference energies are obtained from high-level theory, CCSD(T), while the interactions of reaction intermediates with catalyst surfaces are extracted from computationally less expensive DFT. The approach is demonstrated for the case of graphene-based surfaces as model catalysts in Li-O$_2$ and Na-O$_2$ cells using the minimum viable mechanism. B-doped graphene was identified as the best catalyst among considered surfaces, while pristine graphene performs poorly. Moreover, we show that the inclusion of dispersion corrections for DFT has a significant impact on calculated discharge and charge potentials and suggests that long-range dispersion interactions should always be considered when graphene-based materials are modelled as electrocatalysts. Finally, we offer general guidelines for designing new ORR catalysts for M-O$_2$ cells in terms of the optimization of the interactions of catalyst surface with reaction intermediates.

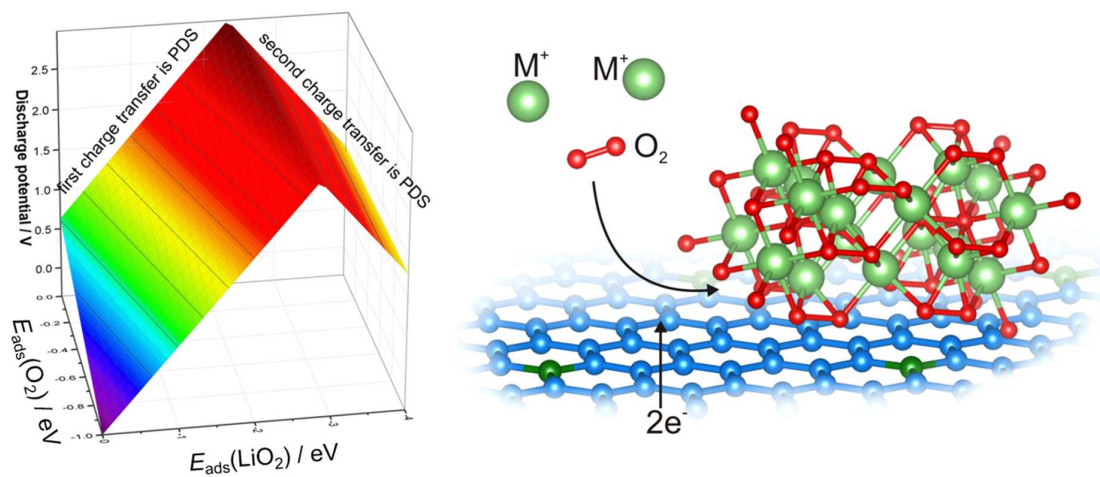

* **Corresponding author:** e-mail: igor@ffh.bg.ac.rs

## 1. Introduction

Since reported by Abraham and Jiang in 1996 [1], Li-$O_2$ batteries have attracted huge interest due to a high energy density, being comparable to gasoline [2], and a possibility to make secondary, rechargeable, cells [3]. With these benefits, Li-$O_2$ batteries represent a suitable alternative to widely used Li-ion batteries and have a perspective for versatile use, including implementation in electric vehicles. In spite of these attractive properties, there are many technical problems which are opposing the practical development of Li-$O_2$ batteries. The most influential of them are limited cycle life, poor electrolyte stability and a decrease in electrical efficiency over continuous cycling [3–6]. Nevertheless, the key prerequisite in developing their practical application is a fundamental understanding of electrode reactions. Moreover, due to a wide usage of Li-ion batteries, the consumption of lithium is a global problem. In a series of reports [7,8], lithium was replaced with sodium and Na-$O_2$ batteries were investigated as a possible alternative for Li-$O_2$ batteries. While the theoretical energy density of Na-$O_2$ batteries is lower than that of Li-$O_2$ batteries [9], Na is much more abundant and cheaper than Li, and the energy density of Na-$O_2$ cells is still much higher than the energy density of common Li-ion cells.

Apart from their advantages and possible applications, Li-$O_2$ and Na-$O_2$ batteries suffer from sluggish kinetics of oxygen reduction reaction (ORR) and oxygen evolution reaction (OER) at the positive electrode, during discharge and charge cycle, respectively. In commercial metal-air cells, like Zn-air cells, electrocatalysts at the air cathode are typically carbon-based materials. However, in the case of Li-$O_2$ cells many other types of electrocatalysts, besides carbons, have been investigated, including metal oxides [10,11] and metals [12]. Typically, platinum is the best catalyst for ORR in aqueous media, and a lot of Pt-based catalysts were also investigated for Li-$O_2$ cells [13,14]. Among possible carbon-based catalysts, materials from the graphene family have attracted a lot of attention due to various benefits. Because of exceptional electrical, mechanical and thermal properties [15], high conductivity and a large surface area [16], graphene has wide application in energy storage [17]. However, pristine graphene is chemically very inert, which hinders its practical application in batteries. On the other hand, graphene functionalized with heteroatoms like boron and nitrogen has improved electrochemical properties and found applications in batteries, including metal-air cells [18–21].

Besides extensive experimental work, theoretical modelling of processes in Li-$O_2$ and Na-$O_2$ cells provides valuable insights into the crucial steps of ORR and OER in the presence of Li$^+$ or Na$^+$ in aprotic media. Most of the existing work addressing elementary steps of ORR and OER relies on the modification of the theoretical ORR model in aqueous media presented by the Nørskov group [22], employing a computational hydrogen electrode approach. When adapted to non-aqueous Li$^+$ containing media, protons are replaced by Li$^+$ while the hydrogen electrode is replaced by Li$^+$/Li electrode [23].

Rather instructive works can be found in the literature, addressing graphene-based materials, metals, and alloys [14,18,24,25]. However, it must be noted that ORR in Li-O$_2$ cells is much more complex than ORR in aqueous media. In the first case, the reaction is 2Li$_{(s)}$ + O$_2$ = Li$_2$O$_{2\,(s)}$, where solid Li$_2$O$_2$ formed at the positive electrode during discharge blocks the electrode surface. In contrast, in aqueous (acidic) media H$_2$O is formed, which does not remain on the surface. Hence, modelling of ORR in Li-O$_2$ cells is much more difficult as one should account for the solid product formation on the electrode surface. Existing reports typically take some complex form of Li$_2$O$_2$ at the catalyst surface as the final product, with various intermediates [24,26,27]. As different surfaces interact differently with these species, the result is that different catalysts considered within the same model give the different electromotive force of hypothetical Li-O$_2$ cell. In other words, *catalyst alters the thermodynamics of the process*, which is in contradiction to the very definition of a catalyst. As mentioned before, this is never the case when modelling ORR in aqueous media as the final product does not remain attached to the catalyst surface, although it is inevitable that the calculated electromotive force differs from the experimental one. In practice, mentioned problems in alteration of thermodynamics and different models of ORR in non-aqueous metal-air cells make a direct comparison of different catalyst rather difficult, if not impossible, even within the same report. As the electromotive force of the cell varies, one is never clear whether calculated discharge potentials or overvoltages should be compared. The same holds for the charging process.

In connection with the identified problems related to the modelling of ORR catalysts for M-O$_2$ cells, our aim is to set the simplest model of ORR applicable for M-O$_2$ cells, which alleviates the influence of catalyst on the thermodynamics of the cell. The model is further applied to investigate electrocatalytic trends in Li-O$_2$ and Na-O$_2$ cells with (doped) graphene cathode. In order to additionally demonstrate the importance of the computational approach, we use Density Functional Theory (DFT) calculations in conjunction with corrections for long-range dispersion interactions.

## 2. Computational details

The first-principles DFT calculations were performed using the Vienna *ab initio* simulation code (VASP) [28–31]. In the first step, we used the generalized gradient approximation (GGA) in the parametrization by Perdew, Burk and Ernzerhof [32] and the projector augmented wave (PAW) method [33,34]. The cut-off energy of 600 eV and Gaussian smearing with a width of $\sigma$ = 0.025 eV for the occupation of the electronic levels were used. A Monkhorst-Pack $\Gamma$-centered 10×10×1 k-point mesh was employed. Adsorption sites for the species involved in charge/discharge processes in Li-O$_2$ and Na-O$_2$ cells were investigated systematically and the most stable sites were considered in the mechanisms presented in this work. The relaxation of all the atoms in the simulation cell was unrestricted. The relaxation procedure was performed until the Hellmann-Feynman forces on all atoms were below $10^{-2}$ eV Å$^{-1}$. Spin-polarization was considered in all of the presented calculations. To account for dispersion interactions, we used DFT theory plus long-range dispersion correction in the DFT+D2 and DFT+D3 formulations of Grimme [35,36]. VESTA code was used for visualization [37]. Reference energies for the isolated species used in this work were obtained at CCSD(T)/aug-cc-pVQZ [38–40] level using Gaussian program package [41]. Frequency calculations confirmed that the optimized structures correspond to stable molecules. All species represent open-shell systems, with an oxygen atom and oxygen molecule being in the triplet state, whereas alkali atoms and their oxides are in the doublet state.

Pristine graphene (*pr*-graphene) was modelled as a single layer with 32 carbon atoms arranged in a honeycomb lattice ($C_{32}$) within the 4×4 supercell and 20 Å of vacuum along the *z*-axis. Doped graphene was modelled by replacing one carbon atom with nitrogen (N-graphene, $C_{31}N$) or boron (B-graphene, $C_{31}B$). In ref. [42] details regarding the electronic structure of such a modified graphene surface can be found. The interactions of species involved in charge/discharge processes at studied surfaces are quantified using adsorption energy, $E_{ads}(A)$:

$$E_{ads}(A) = E_{S+A} - (E_S + E_{A,isol})$$

where $E_{S+A}$, $E_S$, and $E_{A,isol}$ stand for the total energy of substrate with adsorbate A, the total energy of a bare substrate (*pr*-graphene, N-graphene, or B-graphene surface) and the total energy of isolated adsorbate A, respectively. As defined, a lower (more negative) $E_{ads}(A)$ indicates stronger binding. Entropy, zero-point energy and thermal contributions of adsorbed species have been neglected as separate calculations have shown that they contribute less than 0.02 eV to the adsorption energy, in agreement with previous reports, *e.g.* ref. [18] and references therein. Bader algorithm [43,44] was used for the analysis of charge redistribution upon adsorbate-substrate interactions.

## 3. Results and discussion

### 3.1. General assumptions of the model

In order to model the discharge process in Li-$O_2$ and Na-$O_2$ batteries, we consider the following minimal mechanism (where * denotes adsorbed species):

(i)     $M_{(s)} \leftrightarrows M^+ + e^-$                           (M electrode in equilibrium)

(ii)    $O_{2(g)} + 2M^+ + 2e^- \rightarrow O_2^* + 2M^+ + 2e^-$        (air cathode, step 1)

(iii)   $O_2^* + 2M^+ + 2e^- \rightarrow MO_2^* + M^+ + e^-$            (air cathode, step 2)

(iv)    $M^+ + MO_2^* + e^- \rightarrow M_2O_{2(s)}$                   (air cathode, step 3)

The first step on the oxygen cathode ((ii) air cathode, step 1) is the adsorption of molecular oxygen from the atmosphere (solution). In the next step ((iii) air cathode, step 2) the adsorbed $O_2$ reacts with ($M^+$ + $e^-$) to form adsorbed $MO_2$ (first charge transfer). The following step ((iv) air cathode, step 3) is the second metalation (charge transfer) when adsorbed $MO_2$ reacts with ($M^+$ + $e^-$) to form the final product, $M_2O_2$, being incorporated into the lattice of solid $M_2O_2$. The mechanism is represented schematically in **Fig. 1**.

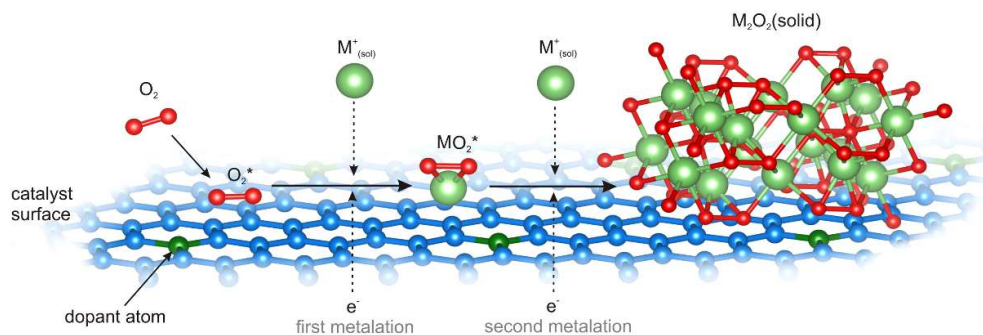

**Figure 1.** Scheme of the basic mechanism ORR in an aprotic solution containing $M^+$. First, $O_2$ is attached to the surface, and the first charge transfer takes place to form adsorbed $MO_2$. With the second charge transfer, $MO_2$ receives one more $M^+$ and gets incorporated into the $M_2O_2$ (solid) lattice.

Obviously, the first intermediate in the proposed mechanism is $O_2^*$. Its adsorption energy is calculated with respect to the isolated $O_2$ molecule, as was defined in Section 2. DFT is known to significantly overestimate the bond energy in $O_2$, resulting in the faulty total energy of isolated $O_2$ molecule, and consequently, unreliable $O_2$ adsorption energies. This problem is usually bypassed by adjusting the energy of isolated $O_2$ using the energetics of the reaction $2H_2 + O_2 = 2H_2O$ [22,23] since $H_2$ and $H_2O$ are reliably described by DFT. Such a procedure reduces the energy of isolated $O_2$ and therefore raises the $O_2$ adsorption energy (*i.e.* suggests weaker binding of $O_2$ to the substrate). In case of $O_2$ on graphene, the DFT calculations indicate weak physisorption [42,45,46], consistent with experimentally determined $O_2$ adsorption energy on graphene of about −0.15 eV [47] (also consistent with the DFT results obtained in the present work, as will be shown later on, in **Table 1**). Therefore, using the aforementioned procedure [22,23] would result in large positive adsorption energy of $O_2$ on graphene (significantly exothermic step (ii) in the given mechanism), which would be in contradiction with the experiment [47]. We note that most of the reports which use the modified scheme of Nørskov *et al.* [22,23] do not consider the $O_2$ adsorption step in the mechanism. In order to avoid both the DFT-$O_2$ problem and the corrective procedure which yields inconsistency with the experiment, in this contribution we calculate the energies of isolated species involved in the mechanism at a higher level of theory,

CCSD(T), in which the $O_2$ molecule is properly described. Then, for the case when these species interact with the surface we correct the chemical potential of the adsorbed species with the adsorption energy from DFT. As the adsorption energies from DFT are generally valid (even for the case of $O_2$ adsorption on graphene), we retain this computationally affordable approach for "large" systems (which include the graphene-based surface), and use the computationally more expensive CCSD(T) scheme only for a few small molecules. Next, in order to obtain the chemical potential of $M^+$, considering that it is in equilibrium with $M_{(s)}$, we take the energy of isolated M atom and add its experimental cohesive energy (negative, 1.63 eV for Li [48,49] and 1.113 eV for Na [48]) to it, assuming that the activity of $M^+$ ions in the solution is 1. Finally, in order to resolve the mentioned problem of electromotive force changing with the catalyst, we consider that the final product is always solid $M_2O_2$ and calculate its free energy using the free energies of $O_2$, M and the experimental value of electromotive force of a given M-$O_2$ cell (2.96 V for Li-$O_2$ [1,3] and 2.33 V for Na-$O_2$ cell [7]). As a result, Gibbs free energies of reactants and the final state are independent of the type of catalyst. Moreover, the energy of $O_2$ molecule is correct (obtained from CCSD(T)), while computational cost of the whole set of calculations is not compromised due to a limited number of computationally expensive calculations.

With the exception of the steps given above, reaction profiles are calculated as in the computational hydrogen electrode approach, that is, its modified Li-case [22,23]. The free energy of electron was assumed to be linearly dependent on the electrode potential and at a different electrode potential $U$ (*vs*. $M^+$/M) has been shifted by $-eU$. The charging process is considered as reversible to discharge (step (iv) → step (ii), solvated $M^+$ remains in equilibrium with solid M). Here we denote equilibrium potential *vs*. $M^+$/M (electromotive force of the cell), discharge and charge potentials as $U_{eq}$, $U_{DC}$ and $U_C$, respectively. As $U_{DC}$ and $U_C$ we take potentials at which there is no thermodynamic barrier for a given process ($\Delta G \leq 0$ for every step) [22].

### 3.2. Adsorption of reactants and intermediates

In order to identify the structures of reactants and intermediates on studied surfaces, we systematically investigated the adsorption of M, $O_2$ and $MO_2$ species, which appear in the reaction mechanism [50–52]. For $MO_2$ we considered isomers to which we refer as triangular (tri) or linear (lin) (see **Fig. 2**). Triangular isomer is the one where M is added to $O_2$ without dissociation of the O−O bond. In contrast, the linear isomer is the one where M is set in the middle of dissociated $O_2$. Energetically more stable isomer on a given surface is taken as the intermediate. We note that the triangular isomer is more stable in the gas phase for both $LiO_2$ and $NaO_2$. Hence, the adsorption energies of $MO_2$ on the studied surfaces are referred to the isolated triangular $MO_2$. In **Table 1** adsorption energies are summarized for all the studied surfaces and used levels of theory.

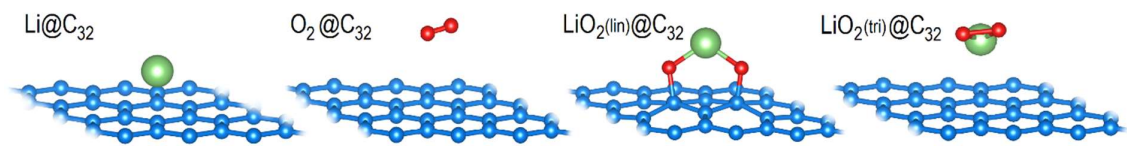

**Figure 2.** Optimized structures of Li-species and $O_2$ adsorbed on pristine graphene.

As noted previously, calculated $O_2$ adsorption energies agree quite well with previous reports, including experimental ones [47]. The adsorption energy of Li is below its cohesive energy (comparing absolute values) [48,49] except in the case of B-doped graphene. The calculated adsorption energies of Li on pristine graphene (Table 1) are in agreement with $E_{ads}$(Li) reported in ref. [53]. The energies of $LiO_2$ intermediates are different depending on the geometry. The reaction will take place *via* $LiO_2$(tri) in cases of pristine graphene and B-doped, while $LiO_2$(lin) is preferred on N-doped graphene. $LiO_2$(lin) on N-doped graphene is more stable than $LiO_2$(tri) by approx. 0.3 eV. The adsorption of Li species on graphene surfaces was followed by the transfer of 0.89 e *per* Li atom (as well as Li from $LiO_2$) from Li atoms to C atoms in the graphene surface, in agreement with previous reports [51,52].

**Table 1.** Adsorption energies (in eV) of studied species (A) on pristine, N-doped and B-doped graphene, calculated using three approaches (PBE, PBE+D2, PBE+D3). For $LiO_2$ and $NaO_2$ (tri) or (lin) indicates preferred isomer configuration on the surface: triangular, or linear.

| A | pristine graphene | | | N-doped graphene | | | B-doped graphene | | |
|---|---|---|---|---|---|---|---|---|---|
| | PBE | PBE+D2 | PBE+D3 | PBE | PBE+D2 | PBE+D3 | PBE | PBE+D2 | PBE+D3 |
| $O_2$ | −0.01 | −0.06 | −0.08 | −0.06 | −0.16 | −0.18 | −0.02 | −0.06 | −0.07 |
| Li | −1.19 | −1.56 | −1.23 | −0.95 | −1.32 | −0.99 | −2.59 | −2.96 | −2.63 |
| Na | −0.55 | −0.93 | −0.64 | −0.23 | −0.66 | −0.37 | −1.78 | −2.22 | −1.93 |
| $LiO_2$ | −0.21 (tri) | −0.57 (tri) | −0.39 (tri) | −0.46 (lin) | −0.84 (lin) | −0.68 (lin) | −0.81 (tri) | −1.30 (tri) | −0.91 (tri) |
| $NaO_2$ | −0.44 (tri) | −0.93 (tri) | −0.65 (tri) | −0.44 (tri) | −0.83 (tri) | −0.65 (lin) | −1.17 (tri) | −1.62 (tri) | −1.31 (tri) |

A comparison between adsorption energies of $LiO_2$ and $NaO_2$ shows that the adsorption is more exothermic for $NaO_2$ on pristine and B-doped graphene, while adsorption energies of $LiO_2$ and $NaO_2$ are found to be similar on N-doped graphene. Interaction between Na and pristine graphene is relatively weak, with Na adsorption energy between −0.55 and −0.93 eV, depending on the computational approach (Table 1). Previous studies reported similar adsorption energies for Na on pristine graphene: −0.35 eV [19], −0.71 eV [19], −0.49 eV [51], −0.462 eV [52] and −0.507 eV [54]. However, the adsorption energy of Na on N-doped graphene is even weaker (by approximately 0.3 eV, Table 1) while in the case of B-doped graphene it is stronger by 1.2-1.3 eV (compared to pristine graphene, Table 1). This is comparable with the work of Yao *et al.* [54] and Dobrota *et al.* [19]. Inclusion of dispersion interactions generally reduces adsorption energies (they become more negative, *i.e.* the bonding is stronger, Table 1). This trend is quite expected. By analyzing the obtained results for the energy of $NaO_2$ in both geometries, we conclude that in the case of pristine and B-doped graphene reaction will occur through $NaO_2$ in (tri) form. However, the adsorption energies of (lin) and (tri) $NaO_2$ are very similar on N-doped graphene and preferred isomer actually depends on the used computational scheme (Table 1). Also, the adsorption of Na species on the studied graphene surfaces was followed by the transfer of 0.7 e *per* Na atom (as well as Na from $NaO_2$ molecules) from Na atoms to C atoms from graphene surface. Transfer of 0.80 e [51] and 0.73 e [52] from Na atom to pristine graphene was reported previously. The analysis of densities of states (DOS, Fig. 3) shows quite similar behavior of different surfaces interacting with $LiO_2$ and $NaO_2$, without alarming band gap opening.

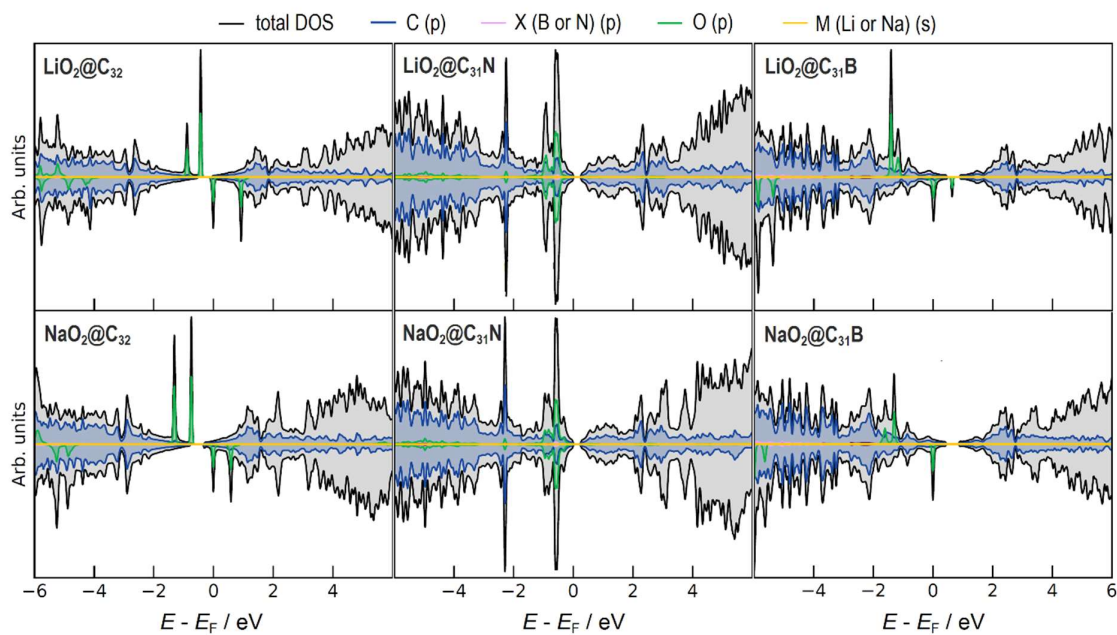

**Figure 3.** DOS plots for pristine (left), N-doped (middle) and B-doped (right) graphene interacting with $LiO_2$ (top row) and $NaO_2$ (bottom row). Plots were made using SUMO tools [55].

### 3.3. Reaction profiles for Li-$O_2$ and Na-$O_2$ battery

For the case of Li-$O_2$ cell we consider the following net reaction [1,3]:

$$2Li + O_2 \rightarrow Li_2O_2$$

so, the only discharge product is $Li_2O_2$. In the case of Na-$O_2$, both $NaO_2$ and $Na_2O_2$ were described as discharge products [56–58]. Because of similar equilibrium potentials for the formation of $NaO_2$ (2.27 V vs. Na+/Na) and Na$_2$O$_2$ (2.33 V vs. Na+/Na), there is a number of experimental and theoretical reports about discharge product in Na-O$_2$ batteries [7–9,57–59]. However, to have a direct comparison with Li-O$_2$ cell we consider the overall reaction:

$$2Na + O_2 \rightarrow Na_2O_2$$

So, there is an equivalency between Li-O$_2$ and Na-O$_2$ systems, and the differences in reaction profiles come from different open-circuit voltages and the strength of the interaction between MO$_2$ intermediates with model catalyst surfaces. First, we consider the effect of the applied computational scheme (*i.e.* inclusion of dispersion interactions). Reference energies are obtained using CCSD(T) as described above. **Fig. 4** gives calculated reaction profiles for pristine graphene for the three calculation schemes we applied (PBE, PBE+D2, and PBE+D3).

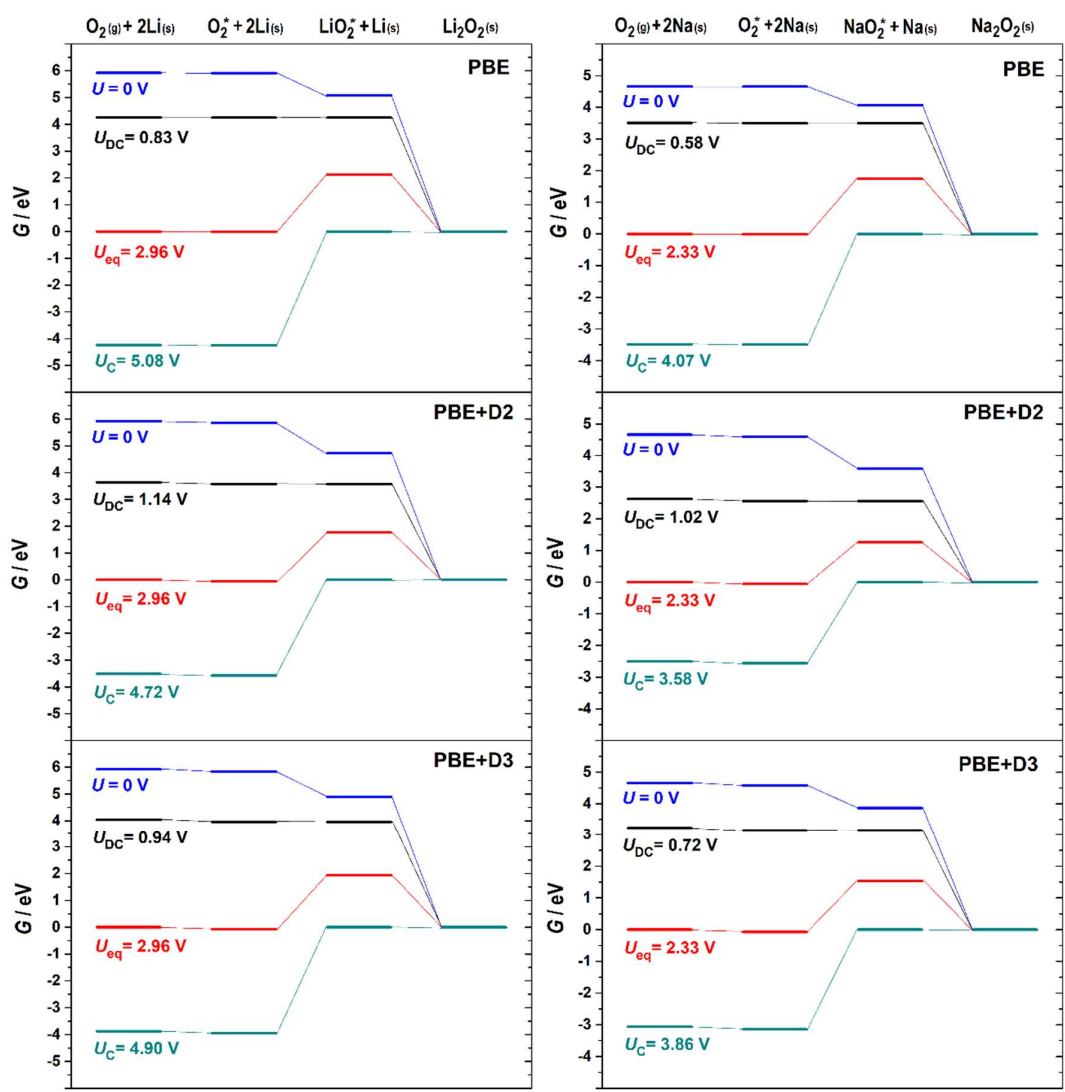

**Figure 4.** Reaction profiles for the case of Li-O$_2$ cell (left) and Na-O$_2$ cell (right) with pristine graphene as the cathode catalyst, calculated using PBE (top row), PBE+D2 (middle row) and PBE+D3 (bottom row).

As can be seen, pristine graphene presents itself as a rather poor catalyst for both Li-O$_2$ and Na-O$_2$ cells. As we consider O$_2$ adsorption as the first step in the process, small negative adsorption energies of O$_2$ give downhill step versus the initial state during discharge. We note that this translates into an uphill step during charge, but with a small barrier (amounting –$E_{ads}$(O$_2$), **Table 1**) that can be overcome at room temperature. So, the discharge potential is determined by the barrier corresponding to the first charge (and M$^+$) transfer. When the reaction is reversed, charge potential is determined by the first oxidation of M$_2$O$_2$ and the formation of MO$_2$*. This is a general feature for all of the considered reaction profiles. Generally, the interactions of MO$_2$* are stronger with the addition of dispersion interactions, PBE+D2 giving stronger bonding than PBE+D3, so MO$_2$* is more stabilized giving higher discharge potential and lower charge potential (**Fig. 4**). An important message comes from these results. As graphene is generally chemically inert and PBE does not account properly for the dispersion interactions, this effect should be taken with great care. In other words, theoretical analysis of the catalytic process on graphene should always account for the long-range dispersion interactions. For example, in the Na-O$_2$ case, the D2 correction nearly doubles $U_{DC}$ on pristine graphene (**Fig. 4**, right).

When considering the effects of different surfaces (**Fig. 5**), we see that N-doped graphene performs only slightly better than pristine graphene. This is in line with our previous results, showing that substitutional doping of graphene with nitrogen does not have a large beneficial effect on the reactivity of the surface. In fact, intermediately formed LiO$_2$ and NaO$_2$ both bind far from the N-dopant site so it is not surprising that pronounced catalytic effect is absent. The effect of N dopant, in this case, is rather indirect, through the disruption of the electronic structure of the graphene basal plane (*n*-doping). The surface with the highest catalytic activity for ORR (and OER) is B-doped graphene, in line with previous reports [18]. However, a high overvoltage of approx. ±1 V (PBE+D2) is needed for both ORR and OER (**Fig. 5**).

We see that in the used approach charge and discharge potentials are quite symmetrical with respect to the equilibrium potential, as the step (ii) in the considered mechanism involves rather small adsorption energies of O$_2$ on considered surfaces. Actually, symmetry is broken by $E_{ads}$(O$_2$) and it is the most obvious when comparing different computational schemes for B-doped graphene (**Fig. 5**) This issue is more elaborated in the work by Hummelshoj *et al.* [27] who reported pronounced asymmetrical charging/discharging potentials, due to different barriers which depend on the number of intermediates considered in the mechanism. An important note is that the first adsorption of O$_2$ is, as a rule, disregarded in the mechanisms considered in the literature. A possible reason, but not a justification, is the fact that this is a great complication. This process does not involve charge transfer and if the adsorption energy of O$_2$ is too large (negative), there will always be a large barrier for charging. If there is positive adsorption energy there will always be a barrier for discharging, as the variation of the electrode potential does not change ΔG for step (ii) in the considered mechanism. Of course, it can be argued that the charge (and M$^+$) transfer is simultaneous with O$_2$ adsorption and that MO$_2$* is formed in this single step. However, our experience from the laboratory says that with the introduction of O$_2$ in de-aerated aprotic electrolytic solution cathode potential increases under open-circuit conditions. This suggests that dissolved O$_2$ interacts with the catalyst surface even without (M$^+$ + e$^-$) transfer.

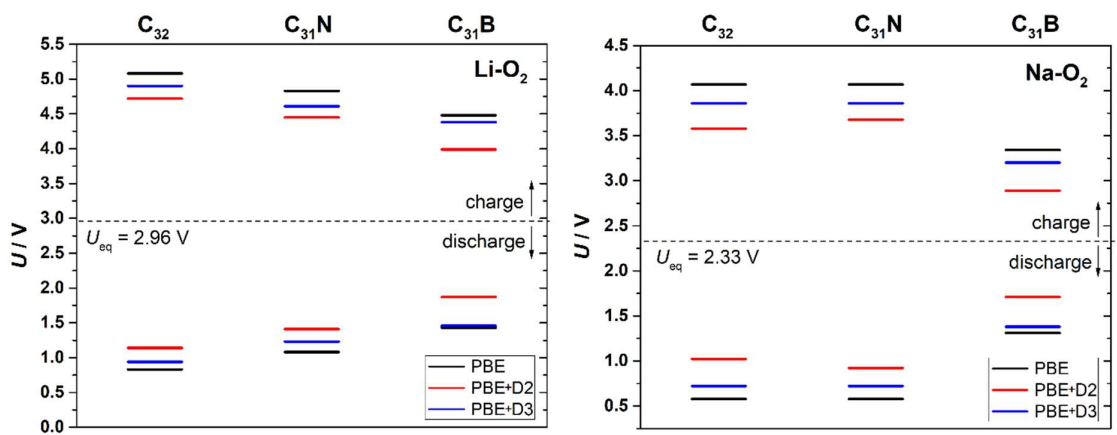

**Figure 5.** Calculated charge and discharge potentials for Li-O$_2$ (left) and Na-O$_2$ cell (right) with pristine, N-doped and B-doped graphene as positive electrode catalysts. Potentials were calculated using PBE, PBE-D2 and PBE-D3 scheme with reference energies from CCSD(T).

As we use a new scheme to construct reaction profiles there is a question of how much this approach affects the final result. So, we constructed reaction profiles and calculated all the discharge and charge potentials using only the energies from DFT. The energy of O$_2$ was used as obtained from the DFT calculations, without the usual type of previously described correction, to have consistent adsorption energies. The energy of bulk M$_2$O$_2$ was then calculated from the experimental open-circuit voltages of Li-O$_2$ or Na-O$_2$ cells. Reaction profiles (**Figs. S1** and **S2**, Supplementary Information) are qualitatively similar to those presented in **Fig. 4**. The same potential determining steps can be identified, and the trends are the same. So, both approaches agree that pristine graphene is the worst considered catalysts, while B-doped graphene is the best. However, the differences are quantitative, in terms of the calculated discharge and charge voltages (**Figs. S3** and **S4**). Namely, when only the DFT energies are used, the calculated values of $U_{DC}$ are much higher and $U_C$ is significantly lower. For pristine, N-doped and B-doped graphene PBE discharge potentials, using only DFT and no dispersion correction, are 1.52 V, 1.76 V and 2.11 V for Li-O$_2$ cell, which is significantly higher than the values obtained using CCSD(T) reference energies (0.83, 1.08 and 1.43 V for the given cases). The constant difference between calculated potentials is due to relative differences in DFT and CCSD(T) energies of LiO$_2$ in the gas phase, DFT giving more stable solution (*i.e.* more exergonic first metalation of O$_2$) reducing in this way thermodynamic barrier corresponding to the first charge transfer step (**Fig. 1**). This is not unexpected as the LiO$_2$ molecule in the gas phase has almost preserved O-O bond whose strength is overestimated by DFT. The situation is the same with NaO$_2$. In conclusion, both schemes give the same trends which are also in line with previous reports, but the exact $U_{DC}$ and $U_C$ are under a great influence of the computational approach used to set the reference energies. The inclusion of dispersion interactions also affects calculated $U_{DC}$ and $U_C$, but the effect is much lower, 10-20% of the calculated potential.

### 3.4. General considerations

The complexity of the ORR process in aprotic media, in the presence of M$^+$, makes its full computational analysis very difficult, if not impossible. Obviously, relative (re)activity trends are the main result of the above-presented analysis, and these are weakly affected by the choice of computational scheme. However, the question is whether a more general consideration can be made. As in the proposed mechanism we have only two intermediates (O$_2$* and MO$_2$*), it is possible to map reactivity in terms of O$_2$ and MO$_2$ adsorption energies on the catalyst surface. Such maps are given in **Fig. 6** for the case when CCSD(T) reference energies are used. **Fig. S5**, Supplementary information, gives analogous maps but when DFT energies are used for the isolated intermediates.

The activity maps can be split into two parts (areas A and B in **Fig. 6**): the one where first charge transfer is the potential determining step and the other one where the second charge transfer determines the discharge potential. For the 3D view see **Fig. S6**, Supplementary information. The first part (A) corresponds to the surfaces which bind MO$_2$ relatively weakly. The second part (B) corresponds to those which bind MO$_2$ strongly so that MO$_2$* tends to stay on the surface instead of getting incorporated in the M$_2$O$_2$ lattice. Also, there is an additional effect of the O$_2$ adsorption energy, which will be discussed later on. The maximum catalytic activities (the highest discharge voltages) are obtained for the surfaces which bind O$_2$ weakly. If the maps are cut with a plane corresponding to $E_{ads}(O_2) = 0$, volcano plots would be obtained, with a single descriptor of ORR activity – the adsorption energy of MO$_2$. Some examples of the correlation between the catalytic activity and the adsorption energies of intermediates for Li-O$_2$ cells can be found in ref. [18], showing generally linear relations. This actually means that all the surfaces studied there lie on the same side of the volcano curve. Using $U_{DC}$ heat maps shown in **Fig. 6**, we find that for Li-O$_2$ case the ideal catalyst should have $E_{ads}(O_2) = 0$ and $E_{ads}(LiO_2) = -2.33$ eV. For the Na-O$_2$ cell the values are $E_{ads}(O_2) = 0$ and $E_{ads}(NaO_2) = -2.18$ eV (for the case of maps designed using only DFT energies the values are $E_{ads}(LiO_2) = -2.13$ eV and $E_{ads}(NaO_2) = -2.46$ eV). So, the recipe for making the perfect catalyst for an M-O$_2$ cell would be to design a material with proper MO$_2$ energetics. Graphene-based surfaces considered here have "good" O$_2$ adsorption energies (*i.e.* close to 0), but they bind MO$_2$ too weakly. Hence, the task to improve a graphene-based catalyst would be to make it bind MO$_2$ more strongly but to maintain weak interaction with O$_2$. However, it is very difficult to imagine a surface that basically does not interact with O$_2$ while binds MO$_2$ with appreciable strength.

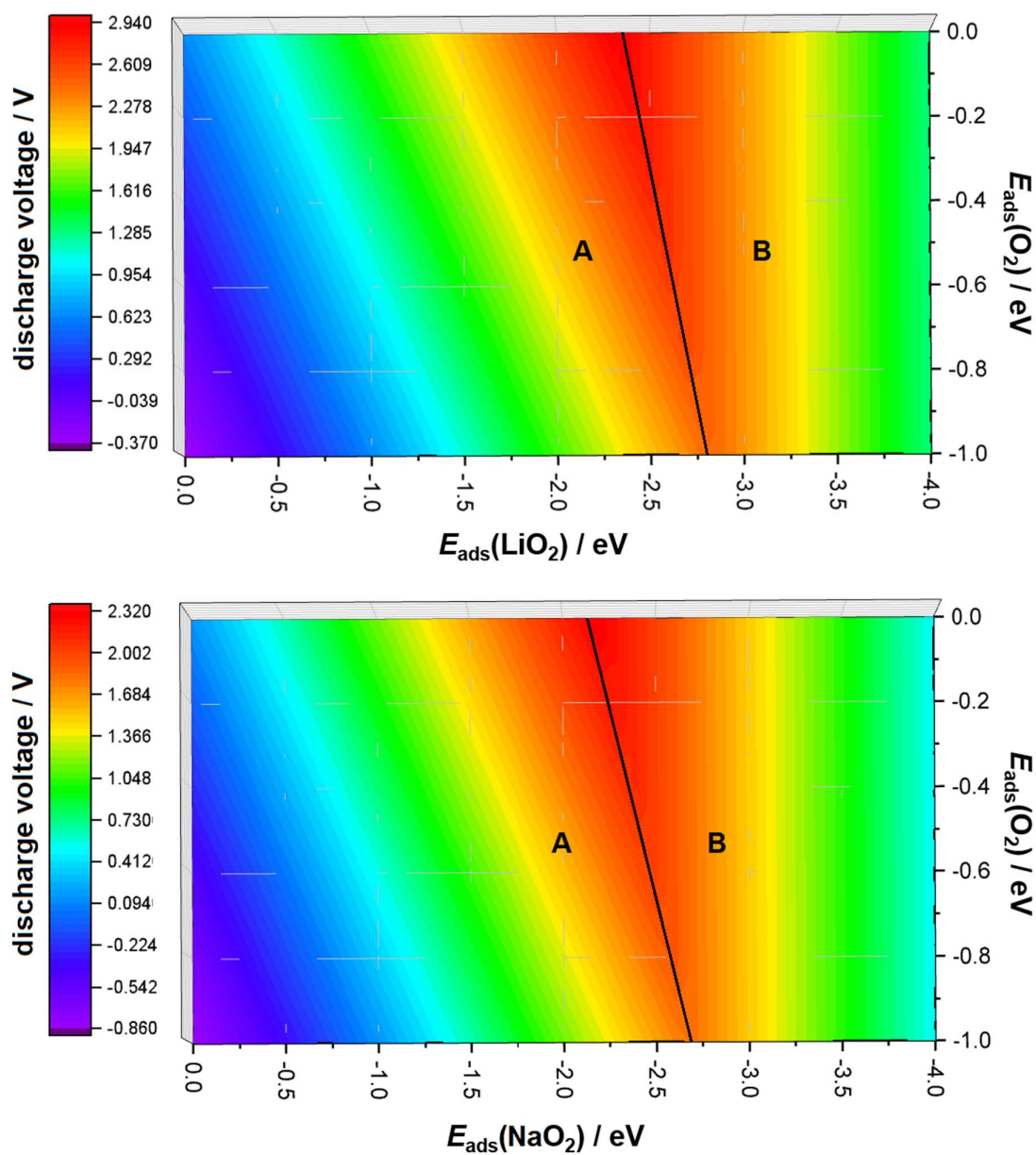

**Figure 6.** Discharge voltage (discharge potential *vs.* M$^+$/M) maps for Li-O$_2$ (top) and Na-O$_2$ (bottom) cell depending on the MO$_2$ and O$_2$ adsorption energies, calculated with CCSD(T) reference energies. The black line on the maps separates areas with different potential determining steps (PDS): area A, where the first charge transfer is PDS and area B, where the second charge transfer is PDS.

In fact, one can expect that certain scaling of $O_2$ and $MO_2$ adsorption energies exists, which is quite common in catalysis [60,61]. If such scaling exists, there would be a cut through the activity map which would also result in a volcano curve and single activity descriptor would, in principle, be enough. However, the question is where this scaling line would go through the activity map? If we consider Pt, it has $O_2$ adsorption energy around −0.5 eV [62]. This sets it already quite far from the "hot" part of the activity map, regardless of the adsorption energy of $MO_2$. If the adsorption energy of $MO_2$ is such that it falls in the part where the first charge transfer is the potential determining step some well know strategies like alloying and compression of the surface [63–66] would do little for the activity as the $MO_2$ adsorption energy would also be weakened. One more important question is whether there is scaling between the adsorption energies of $LiO_x$ and corresponding $NaO_x$ species. This would allow one to merge the activity maps for Li-$O_2$ and Na-$O_2$ cells while obtaining a single (general) descriptor for designing new catalysts for M-$O_2$ cells. For example, let us look at the mechanism discussed in this contribution: if there is a correlation between $LiO_2$ and $NaO_2$ adsorption energies, these cells would not have to be considered separately. By knowing how one adsorbate binds to a given surface, the binding of the other one would be known as well. In that way, only one descriptor (adsorption energy of one $MO_2$ specie) would be needed. However, we expect that different classes of materials (carbon materials, metals, oxides, etc.) would populate different parts of activity map(s) as, in general, there is no reason to expect that all these materials follow a unique scaling relation. We have checked for this possibility by plotting the adsorption energy of $NaO_2$ *vs.* adsorption energy of $LiO_2$ on corresponding studied surfaces (Fig. 7). We find that for each of the investigated surfaces the datapoints obtained using the three mentioned approaches (PBE, PBE+D2 and PBE+D3) correlate linearly. Additionally, the points for $MO_2$ adsorption on pristine and B-doped graphene seem to follow a similar linear relationship, while the N-doped case is shifted towards more positive $NaO_2$ adsorption energies. Obviously, the difference between the materials does not have to be so significant (as previously mentioned) – the type of the dopant atom can be enough to induce adsorption energy scaling changes.

Another important conclusion that can be deduced from the 3D view presented in Fig. S6 is the one that the ORR activity "responds" differently to the changes of $E_{ads}(O_2)$ and $E_{ads}(MO_2)$. Activity is much more sensitive to the change of $E_{ads}(MO_2)$, particularly in the strongly binding branch of the activity map, where $O_2$ adsorption energy does not affect the discharge potential. This means that on this side of the volcano, the only task would be to weaken catalyst interaction with $MO_2$ and approach its adsorption energies to ideal "target" values, without the need to take great care of the changes of $O_2$ adsorption energy. However, in this case, the maximum achievable discharge potential would depend on $O_2$ adsorption energy.

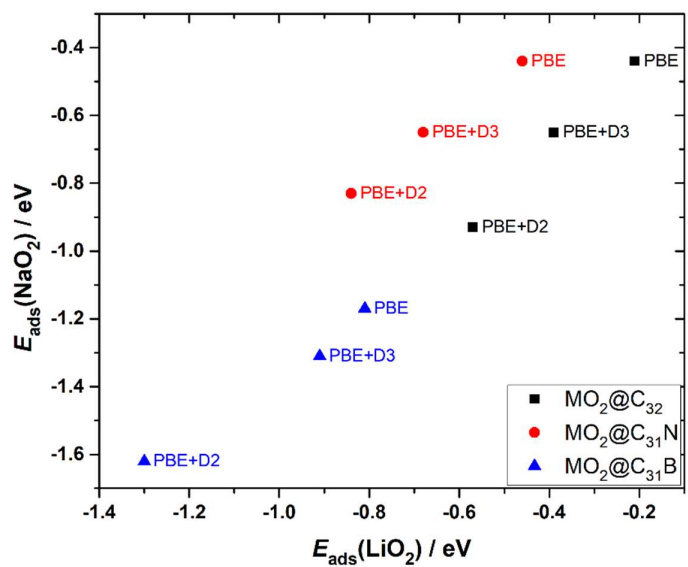

**Figure 7.** Correlation between NaO$_2$ and LiO$_2$ energies of adsorption on C$_{32}$, C$_{31}$N and C$_{31}$B surfaces, obtained using PBE, PBE+D2 and PBE+D3 computational schemes.

## 4. Conclusions

We present the analysis of the impact of the computational approach used to model processes in M-O$_2$ cells. In order to make thermodynamics of the cell independent on the catalyst surface, which has been identified as an issue in the existing reports, we analyze a minimal model for Li-O$_2$ and Na-O$_2$ cells, where high-level CCSD(T) is used to set reference energies in the reaction mechanism, while adsorption energies are obtained from DFT. So, the computational cost of this approach is not significantly affected, while the aforementioned problem is mitigated. In addition, we show that the inclusion of dispersion interactions is of high importance when analyzing graphene-based materials as positive electrode catalysts in M-O$_2$ cells. Considering catalytic activity, pristine graphene is poor ORR and OER catalyst for both Li-O$_2$ and Na-O$_2$ cells, while nitrogen doping improves its catalytic activity only slightly. B-doped graphene is identified as the best catalyst among considered surfaces. In a more general consideration, we devise general strategies to maximize the ORR performance of the catalyst, showing that ORR activity is much more sensitive to the changes of MO$_2$ adsorption energy. So, the task for designing of new ORR catalysts for M-O$_2$ cells would be to optimize MO$_2$ bonding in order to reach optimal values.

Finally, we propose that the approach where high-level theory is used to obtain reference energies for the construction of reaction profiles could be of general use and can be applied for the modelling of a more complex reaction mechanism in M-O$_2$ cells. The inclusion of solid M$_2$O$_2$ (or any other final solid product) is recommended as it alleviates the dependence of cell thermodynamics on the model catalyst surface and allows for a direct comparison of different materials. In fact, we consider that any (electro)catalytic reaction can be treated using this approach, particularly when the standard DFT scheme is not capable to properly describe the ground state of any of the reaction intermediates.

**Acknowledgments**

This research was supported by the Serbian Ministry of Education, Science and Technological Development. This research was sponsored in part by the NATO Science for Peace and Security Programme under grant G5729. S.V.M is indebted to the Serbian Academy of Sciences and Arts for funding the study through the project "Electrocatalysis in the contemporary processes of energy conversion". N.V.S. acknowledges the support provided by the Swedish Research Council through the project no. 2014-5993. We also acknowledge the support from the Carl Tryggers Foundation for Scientific Research (grant no. 18:177), Sweden. The computations were performed on resources provided by the Swedish National Infrastructure for Computing (SNIC) at National Supercomputer Centre (NSC) at Linköping University.

**References**

[1] K.M. Abraham, Z. Jiang, A polymer electrolyte-based rechargeable lithium/oxygen battery, J. Electrochem. Soc. 143 (1996) 1–5. doi:10.1149/1.1836378.

[2] G. Girishkumar, B. McCloskey, A.C. Luntz, S. Swanson, W. Wilcke, Lithium−Air Battery: Promise and Challenges, J. Phys. Chem. Lett. 1 (2010) 2193–2203. doi:10.1021/jz1005384.

[3] T. Ogasawara, A. Débart, M. Holzapfel, P. Novák, P.G. Bruce, Rechargeable $Li_2O_2$ electrode for lithium batteries, J. Am. Chem. Soc. 128 (2006) 1390–1393. doi:10.1021/ja056811q.

[4] R.R. Mitchell, B.M. Gallant, C. V. Thompson, Y. Shao-Horn, All-carbon-nanofiber electrodes for high-energy rechargeable Li-O 2 batteries, Energy Environ. Sci. 4 (2011) 2952–2958. doi:10.1039/c1ee01496j.

[5] B.D. McCloskey, A. Speidel, R. Scheffler, D.C. Miller, V. Viswanathan, J.S. Hummelshøj, J.K. Nørskov, A.C. Luntz, Twin problems of interfacial carbonate formation in nonaqueous Li-O 2 batteries, J. Phys. Chem. Lett. 3 (2012) 997–1001. doi:10.1021/jz300243r.

[6] B.D. McCloskey, D.S. Bethune, R.M. Shelby, G. Girishkumar, A.C. Luntz, Solvents critical role in nonaqueous Lithium-Oxygen battery electrochemistry, J. Phys. Chem. Lett. 2 (2011) 1161–1166. doi:10.1021/jz200352v.

[7]     P. Hartmann, C.L. Bender, M. Vračar, A.K. Dürr, A. Garsuch, J. Janek, P. Adelhelm, A rechargeable room-temperature sodium superoxide (NaO2) battery, Nat. Mater. 12 (2013) 228–232. doi:10.1038/nmat3486.

[8]     E. Peled, D. Golodnitsky, H. Mazor, M. Goor, S. Avshalomov, Parameter analysis of a practical lithium- and sodium-air electric vehicle battery, J. Power Sources. 196 (2011) 6835–6840. doi:10.1016/j.jpowsour.2010.09.104.

[9]     I. Landa-Medrano, C. Li, N. Ortiz-Vitoriano, I. Ruiz de Larramendi, J. Carrasco, T. Rojo, Sodium–Oxygen Battery: Steps Toward Reality, J. Phys. Chem. Lett. 7 (2016) 1161–1166. doi:10.1021/acs.jpclett.5b02845.

[10]    J. Li, N. Wang, Y. Zhao, Y. Ding, L. Guan, MnO2 nanoflakes coated on multi-walled carbon nanotubes for rechargeable lithium-air batteries, Electrochem. Commun. 13 (2011) 698–700. doi:10.1016/j.elecom.2011.04.013.

[11]    A. Débart, A.J. Paterson, J. Bao, P.G. Bruce, α-MnO2 Nanowires: A Catalyst for the O2 Electrode in Rechargeable Lithium Batteries, Angew. Chemie Int. Ed. 47 (2008) 4521–4524. doi:10.1002/anie.200705648.

[12]    J.R. Harding, Y.C. Lu, Y. Tsukada, Y. Shao-Horn, Evidence of catalyzed oxidation of Li 2 O 2 for rechargeable Li-air battery applications, Phys. Chem. Chem. Phys. 14 (2012) 10540–10546. doi:10.1039/c2cp41761h.

[13]    J. Kim, S.W. Lee, C. Carlton, Y. Shao-Horn, Oxygen reduction activity of PtxNi1-x alloy nanoparticles on multiwall carbon nanotubes, Electrochem. Solid-State Lett. 14 (2011) B110. doi:10.1149/1.3613677.

[14]    B.G. Kim, H.J. Kim, S. Back, K.W. Nam, Y. Jung, Y.K. Han, J.W. Choi, Improved reversibility in lithium-oxygen battery: Understanding elementary reactions and surface charge engineering of metal alloy catalyst, Sci. Rep. 4 (2014) 1–9. doi:10.1038/srep04225.

[15]    A.K. Geim, K.S. Novoselov, The rise of graphene, Nat. Mater. 6 (2007) 183–191. doi:10.1038/nmat1849.

[16]    A.K. Geim, Graphene: Status and prospects, Science (80-. ). 324 (2009) 1530–1534. doi:10.1126/science.1158877.

[17]    R. Raccichini, A. Varzi, S. Passerini, B. Scrosati, The role of graphene for electrochemical energy storage, Nat. Mater. 14 (2015) 271–279. doi:10.1038/nmat4170.

[18] H.R. Jiang, T.S. Zhao, L. Shi, P. Tan, L. An, First-Principles Study of Nitrogen-, Boron-Doped Graphene and Co-Doped Graphene as the Potential Catalysts in Nonaqueous Li–$O_2$ Batteries, J. Phys. Chem. C. 120 (2016) 6612–6618. doi:10.1021/acs.jpcc.6b00136.

[19] A.S. Dobrota, I.A. Pašti, S. V. Mentus, B. Johansson, N. V. Skorodumova, Functionalized graphene for sodium battery applications: the DFT insights, Electrochim. Acta. 250 (2017) 185–195. doi:10.1016/j.electacta.2017.07.186.

[20] S.Y. Kim, H.T. Lee, K.B. Kim, Electrochemical properties of graphene flakes as an air cathode material for Li-$O_2$ batteries in an ether-based electrolyte, Phys. Chem. Chem. Phys. 15 (2013) 20262–20271. doi:10.1039/c3cp53534g.

[21] G. Wu, N.H. Mack, W. Gao, S. Ma, R. Zhong, J. Han, J.K. Baldwin, P. Zelenay, Nitrogen-Doped Graphene-Rich Catalysts Derived from Heteroatom Polymers for Oxygen Reduction in Nonaqueous Lithium–$O_2$ Battery Cathodes, ACS Nano. 6 (2012) 9764–9776. doi:10.1021/nn303275d.

[22] J.K. Nørskov, J. Rossmeisl, A. Logadottir, L. Lindqvist, J.R. Kitchin, T. Bligaard, H. Jónsson, Origin of the overpotential for oxygen reduction at a fuel-cell cathode, J. Phys. Chem. B. 108 (2004) 17886–17892. doi:10.1021/jp047349j.

[23] J.S. Hummelshøj, A.C. Luntz, J.K. Nørskov, Theoretical evidence for low kinetic overpotentials in Li-O2 electrochemistry, J. Chem. Phys. 138 (2013) 034703. doi:10.1063/1.4773242.

[24] Y. Xu, W.A. Shelton, O2 reduction by lithium on Au(111) and Pt(111), J. Chem. Phys. 133 (2010) 024703. doi:10.1063/1.3447381.

[25] H.J. Kim, S.C. Jung, Y.K. Han, S.H. Oh, An atomic-level strategy for the design of a low overpotential catalyst for Li-O2 batteries, Nano Energy. 13 (2015) 679–686. doi:10.1016/j.nanoen.2015.03.030.

[26] C.O. Laoire, S. Mukerjee, K.M. Abraham, E.J. Plichta, M.A. Hendrickson, Elucidating the Mechanism of Oxygen Reduction for Lithium-Air Battery Applications, J. Phys. Chem. C. 113 (2009) 20127–20134. doi:10.1021/jp908090s.

[27] J.S. Hummelshøj, J. Blomqvist, S. Datta, T. Vegge, J. Rossmeisl, K.S. Thygesen, A.C. Luntz, K.W. Jacobsen, J.K. Nørskov, Communications: Elementary oxygen electrode reactions in the aprotic Li-air battery, J. Chem. Phys. 132 (2010) 071101. doi:10.1063/1.3298994.

[28] G. Kresse, J. Hafner, Ab initio molecular dynamics for liquid metals, Phys. Rev. B. 47 (1993)

558–561. doi:10.1103/PhysRevB.47.558.

[29] G. Kresse, J. Hafner, Ab initio molecular-dynamics simulation of the liquid-metalamorphous-semiconductor transition in germanium, Phys. Rev. B. 49 (1994) 14251–14269. doi:10.1103/PhysRevB.49.14251.

[30] G. Kresse, J. Furthmüller, Efficiency of ab-initio total energy calculations for metals and semiconductors using a plane-wave basis set, Comput. Mater. Sci. 6 (1996) 15–50. doi:10.1016/0927-0256(96)00008-0.

[31] G. Kresse, J. Furthmüller, Efficient iterative schemes for ab initio total-energy calculations using a plane-wave basis set, Phys. Rev. B. 54 (1996) 11169–11186. doi:10.1103/PhysRevB.54.11169.

[32] J.P. Perdew, K. Burke, M. Ernzerhof, Generalized Gradient Approximation Made Simple, Phys. Rev. Lett. 77 (1996) 3865–3868. doi:10.1103/PhysRevLett.77.3865.

[33] P.E. Blöchl, Projector augmented-wave method, Phys. Rev. B. 50 (1994) 17953–17979. doi:10.1103/PhysRevB.50.17953.

[34] G. Kresse, D. Joubert, From ultrasoft pseudopotentials to the projector augmented-wave method, Phys. Rev. B - Condens. Matter Mater. Phys. 59 (1999) 1758–1775. doi:10.1103/PhysRevB.59.1758.

[35] S. Grimme, Semiempirical GGA-type density functional constructed with a long-range dispersion correction, J. Comput. Chem. 27 (2006) 1787–1799. doi:10.1002/jcc.20495.

[36] S. Grimme, J. Antony, S. Ehrlich, H. Krieg, A consistent and accurate ab initio parametrization of density functional dispersion correction (DFT-D) for the 94 elements H-Pu, J. Chem. Phys. 132 (2010) 154104. doi:10.1063/1.3382344.

[37] K. Momma, F. Izumi, VESTA: A three-dimensional visualization system for electronic and structural analysis, J. Appl. Crystallogr. 41 (2008) 653–658. doi:10.1107/S0021889808012016.

[38] G.D. Purvis, R.J. Bartlett, A full coupled-cluster singles and doubles model: The inclusion of disconnected triples, J. Chem. Phys. 76 (1982) 1910–1918. doi:10.1063/1.443164.

[39] T.H. Dunning, Gaussian basis sets for use in correlated molecular calculations. I. The atoms boron through neon and hydrogen, J. Chem. Phys. 90 (1989) 1007–1023. doi:10.1063/1.456153.

[40] R.A. Kendall, T.H. Dunning, R.J. Harrison, Electron affinities of the first-row atoms revisited. Systematic basis sets and wave functions, J. Chem. Phys. 96 (1992) 6796–6806.

doi:10.1063/1.462569.

[41]   M.J. Frisch, G.W. Trucks, H.B. Schlegel, G.E. Scuseria, M.A. Robb, J.R. Cheeseman, G. Scalmani, V. Barone, B. Mennucci, G.A. Petersson, H. Nakatsuji, M. Caricato, X. Li, H.P. Hratchian, A.F. Izmaylov, J. Bloino, G. Zheng, J.L. Sonnenberg, M. Hada, M. Ehara, K. Toyota, R. Fukuda, J. Hasegawa, M. Ishida, T. Nakajima, Y. Honda, O. Kitao, H. Nakai, T. Vreven, J.A. Montgomery, J.E. Peralta, F. Ogliaro, M. Bearpark, J.J. Heyd, E. Brothers, K.N. Kudin, V.N. Staroverov, R. Kobayashi, J. Normand, K. Raghavachari, A. Rendell, J.C. Burant, S.S. Iyengar, J. Tomasi, M. Cossi, N. Rega, J.M. Millam, M. Klene, J.E. Knox, J.B. Cross, V. Bakken, C. Adamo, J. Jaramillo, R. Gomperts, R.E. Stratmann, O. Yazyev, A.J. Austin, R. Cammi, C. Pomelli, J.W. Ochterski, R.L. Martin, K. Morokuma, V.G. Zakrzewski, G.A. Voth, P. Salvador, J.J. Dannenberg, S. Dapprich, A.D. Daniels, Ö. Farkas, J.B. Foresman, J. V. Ortiz, J. Cioslowski, D.J. Fox, Gaussian 09, Revision D.01, Gaussian, Inc., Wallingford CT, (2009).

[42]   A.S. Dobrota, I.A. Pašti, S. V. Mentus, N. V. Skorodumova, A DFT study of the interplay between dopants and oxygen functional groups over the graphene basal plane – implications in energy-related applications, Phys. Chem. Chem. Phys. 19 (2017) 8530–8540. doi:10.1039/C7CP00344G.

[43]   R. F. M. Bader, R.F.M. Bader, Atoms in molecules. A quantum theory. Oxford:, 1995.

[44]   G. Henkelman, A. Arnaldsson, H. Jónsson, A fast and robust algorithm for Bader decomposition of charge density, Comput. Mater. Sci. 36 (2006) 354–360. doi:10.1016/j.commatsci.2005.04.010.

[45]   H.J. Yan, B. Xu, S.Q. Shi, C.Y. Ouyang, First-principles study of the oxygen adsorption and dissociation on graphene and nitrogen doped graphene for Li-air batteries, J. Appl. Phys. 112 (2012) 104316. doi:10.1063/1.4766919.

[46]   P. Giannozzi, R. Car, G. Scoles, Oxygen adsorption on graphite and nanotubes, J. Chem. Phys. 118 (2003) 1003–1006. doi:10.1063/1.1536636.

[47]   F.R. Bagsican, A. Winchester, S. Ghosh, X. Zhang, L. Ma, M. Wang, H. Murakami, S. Talapatra, R. Vajtai, P.M. Ajayan, J. Kono, M. Tonouchi, I. Kawayama, Adsorption energy of oxygen molecules on graphene and two-dimensional tungsten disulfide, Sci. Rep. 7 (2017) 1–10. doi:10.1038/s41598-017-01883-1.

[48]   Kittel C, Introduction to Solid State Physics, 8th edition, Berkeley, 1996.

[49]   E. Kaxiras, Atomic and electronic structure of solids, Cambridge University Press, 2003.

[50] J.H. Lee, S.G. Kang, H.S. Moon, H. Park, I.T. Kim, S.G. Lee, Adsorption mechanisms of lithium oxides ($Li_xO_2$) on a graphene-based electrode: A density functional theory approach, Appl. Surf. Sci. 351 (2015) 193–202. doi:10.1016/j.apsusc.2015.05.119.

[51] A.S. Dobrota, I.A. Pašti, N. V. Skorodumova, Oxidized graphene as an electrode material for rechargeable metal-ion batteries - a DFT point of view, Electrochim. Acta. 176 (2015) 1092–1099. doi:10.1016/j.electacta.2015.07.125.

[52] K.T. Chan, J.B. Neaton, M.L. Cohen, First-principles study of metal adatom adsorption on graphene, Phys. Rev. B - Condens. Matter Mater. Phys. 77 (2008) 235430. doi:10.1103/PhysRevB.77.235430.

[53] I.A. Pašti, A. Jovanović, A.S. Dobrota, S. V. Mentus, B. Johansson, N. V. Skorodumova, Atomic adsorption on pristine graphene along the Periodic Table of Elements – From PBE to non-local functionals, Appl. Surf. Sci. 436 (2018) 433–440. doi:10.1016/j.apsusc.2017.12.046.

[54] L.-H. Yao, M.-S. Cao, H.-J. Yang, X.-J. Liu, X.-Y. Fang, J. Yuan, Adsorption of Na on intrinsic, B-doped, N-doped and vacancy graphenes: A first-principles study, Comput. Mater. Sci. 85 (2014) 179–185. doi:10.1016/j.commatsci.2013.12.052.

[55] A. M Ganose, A. J Jackson, D. O Scanlon, sumo: Command-line tools for plotting and analysis of periodic ab initio calculations, J. Open Source Softw. 3 (2018) 717. doi:10.21105/joss.00717.

[56] S. Kang, Y. Mo, S.P. Ong, G. Ceder, Nanoscale stabilization of sodium oxides: Implications for Na-O2 batteries, Nano Lett. 14 (2014) 1016–1020. doi:10.1021/nl404557w.

[57] H. Yadegari, Y. Li, M.N. Banis, X. Li, B. Wang, Q. Sun, R. Li, T.K. Sham, X. Cui, X. Sun, On rechargeability and reaction kinetics of sodium-air batteries, Energy Environ. Sci. 7 (2014) 3747–3757. doi:10.1039/c4ee01654h.

[58] Y. Li, H. Yadegari, X. Li, M.N. Banis, R. Li, X. Sun, Superior catalytic activity of nitrogen-doped graphene cathodes for high energy capacity sodium-air batteries, Chem. Commun. 49 (2013) 11731–11733. doi:10.1039/c3cc46606j.

[59] B.D. McCloskey, J.M. Garcia, A.C. Luntz, Chemical and electrochemical differences in nonaqueous Li-O2 and Na-O2 batteries, J. Phys. Chem. Lett. 5 (2014) 1230–1235. doi:10.1021/jz500494s.

[60] F. Abild-Pedersen, J. Greeley, F. Studt, J. Rossmeisl, T.R. Munter, P.G. Moses, E. Skúlason, T. Bligaard, J.K. Nørskov, Scaling Properties of Adsorption Energies for Hydrogen-Containing

Molecules on Transition-Metal Surfaces, Phys. Rev. Lett. 99 (2007) 016105. doi:10.1103/PhysRevLett.99.016105.

[61] A.S. Dobrota, I.A. Pašti, S. V. Mentus, N. V. Skorodumova, A general view on the reactivity of the oxygen-functionalized graphene basal plane, Phys. Chem. Chem. Phys. 18 (2016) 6580–6586. doi:10.1039/c5cp07612a.

[62] A. Groß, Ab initio molecular dynamics simulations of the O2/Pt(1 1 1) interaction, Catal. Today. 260 (2016) 60–65. doi:10.1016/j.cattod.2015.04.040.

[63] I.A. Pašti, E. Fako, A.S. Dobrota, N. López, N. V. Skorodumova, S. V. Mentus, Atomically Thin Metal Films on Foreign Substrates: From Lattice Mismatch to Electrocatalytic Activity, ACS Catal. 9 (2019) 3467–3481. doi:10.1021/acscatal.8b04236.

[64] B. Hammer, J.K. Nørskov, Theoretical surface science and catalysis—calculations and concepts, Adv. Catal. 45 (2000) 71–129. doi:10.1016/S0360-0564(02)45013-4.

[65] E. Fako, A.S. Dobrota, I.A. Pašti, N. López, S. V. Mentus, N. V. Skorodumova, Lattice mismatch as the descriptor of segregation, stability and reactivity of supported thin catalyst films, Phys. Chem. Chem. Phys. 20 (2018) 1524–1530. doi:10.1039/c7cp07276g.

[66] J.R. Kitchin, J.K. Nørskov, M.A. Barteau, J.G. Chen, Modification of the surface electronic and chemical properties of Pt(111) by subsurface 3d transition metals, J. Chem. Phys. 120 (2004) 10240–10246. doi:10.1063/1.1737365.

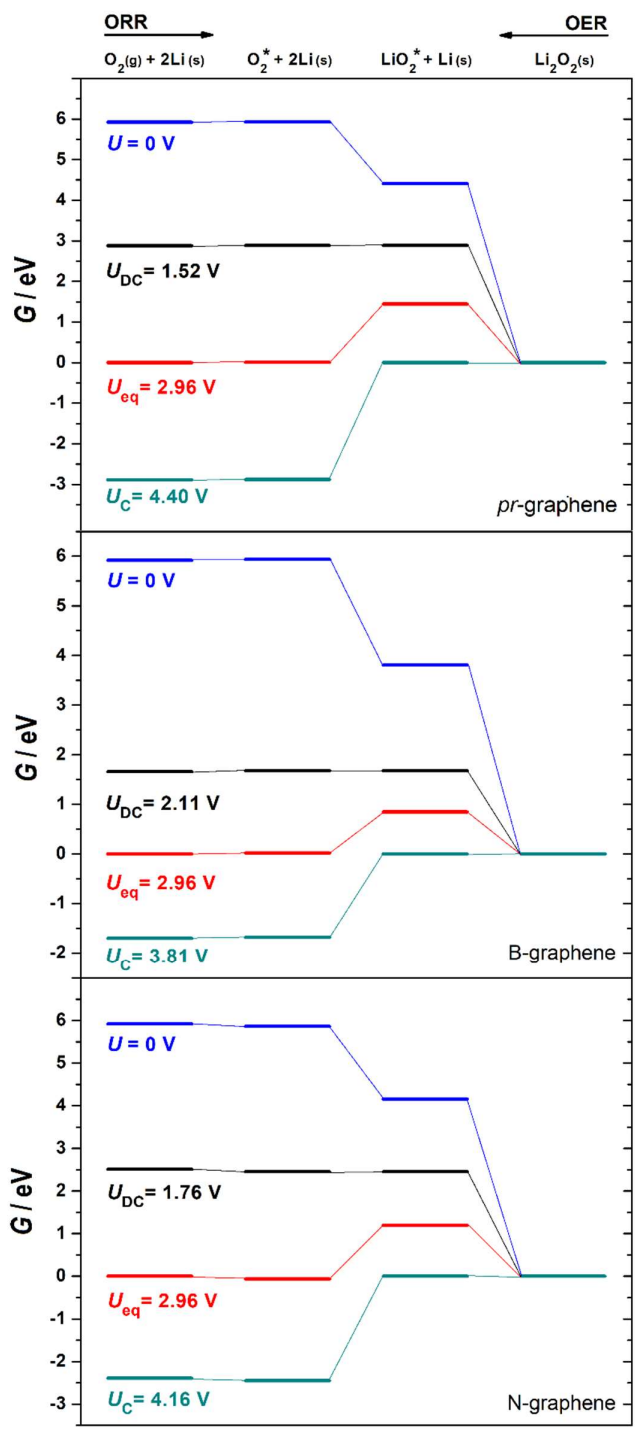

**Figure S1.** Energy profiles for ORR and OER over a) pristine graphene; b) N-graphene and c) B-graphene surfaces, at equilibrium ($U_{eq}$), charging ($U_C$), discharging ($U_{DC}$) potentials and $U$ = 0 V.

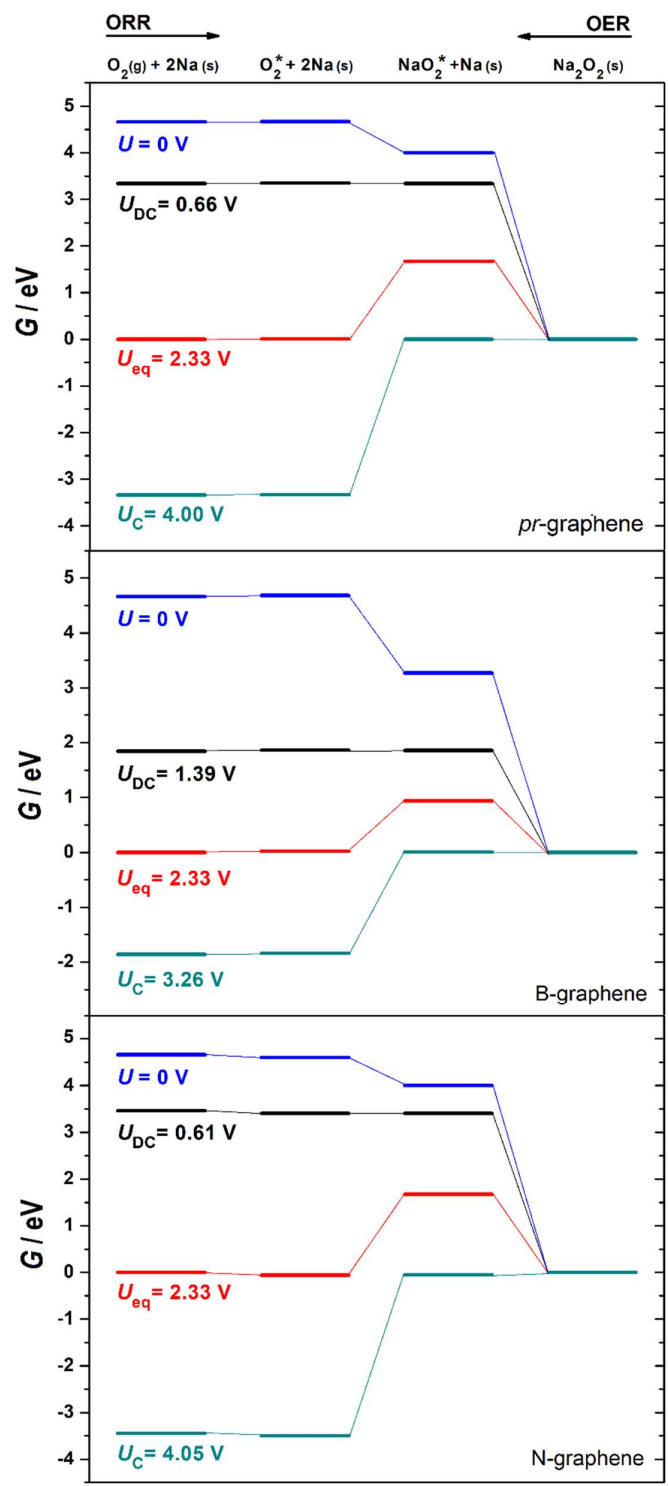

**Figure S2.** Energy diagram for oxygen reduction in a sodium-air cell, over (top) *pr*-graphene; (middle) N-graphene and (bottom) B-graphene surfaces. There is shown equilibrium ($U_{eq}$), charging ($U_C$), discharging ($U_{DC}$) potentials and $U$ = 0V.

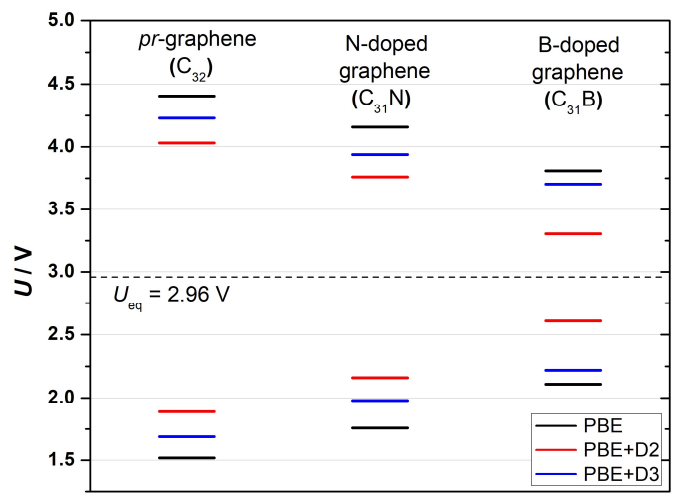

**Figure S3.** Charge and discharge potentials for Li-$O_2$ battery depending on the cathode catalyst and the level of theory; equilibrium potential is indicated by horizontal, dashed line ($U_{eq}$ = 2.96 V).

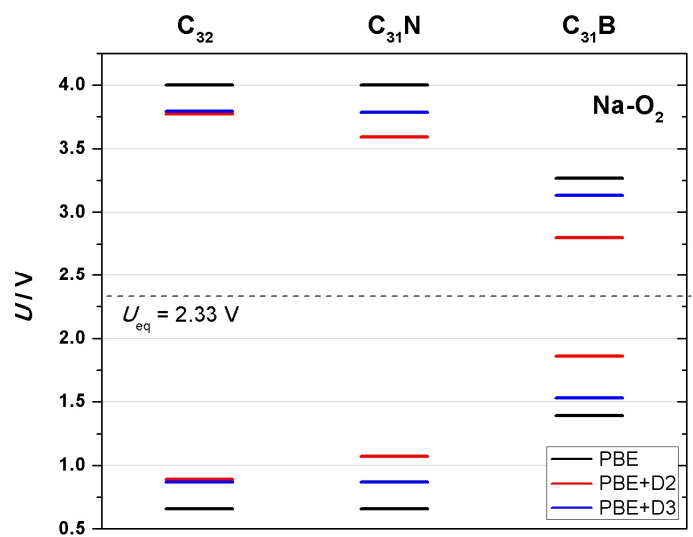

**Figure S4.** Charge and discharge potentials for Na-$O_2$ battery depending on the cathode catalyst and the level of theory; equilibrium potential is indicated by horizontal, dashed line ($U_{eq}$ = 2.33 V).

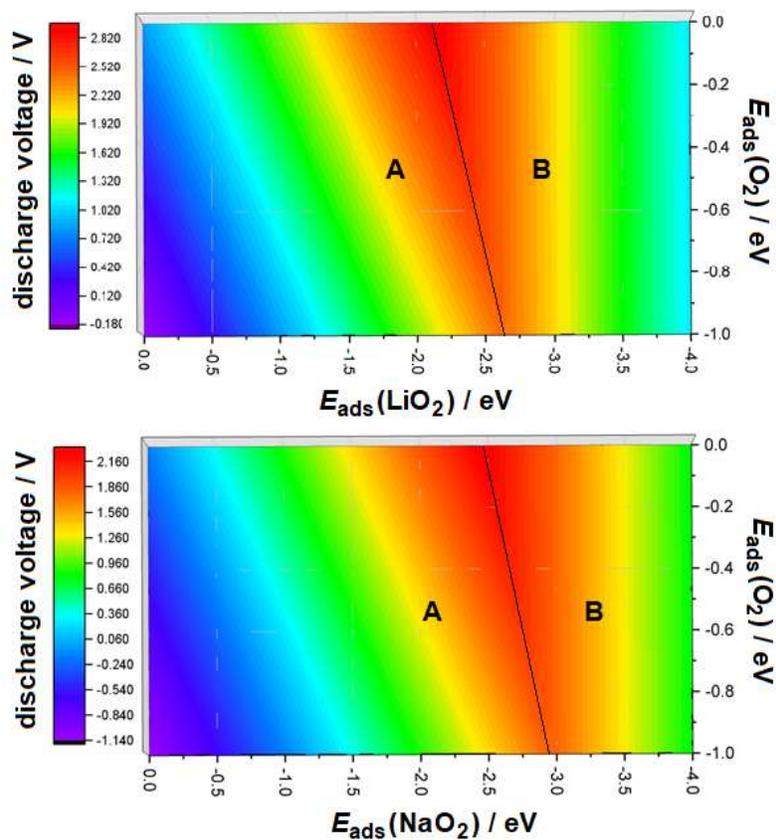

**Figure S5.** Discharge voltage (discharge potential *vs.* M+/M) maps for Li-O$_2$ (top) and Na-O$_2$ (bottom) cell depending on the MO$_2$ and O$_2$ adsorption energies. The black line on the maps separates areas with different potential determining steps (PDS): area A, where the first charge transfer is PDS and area B, where the second charge transfer is PDS. Maps are constructed using the energies from DFT.

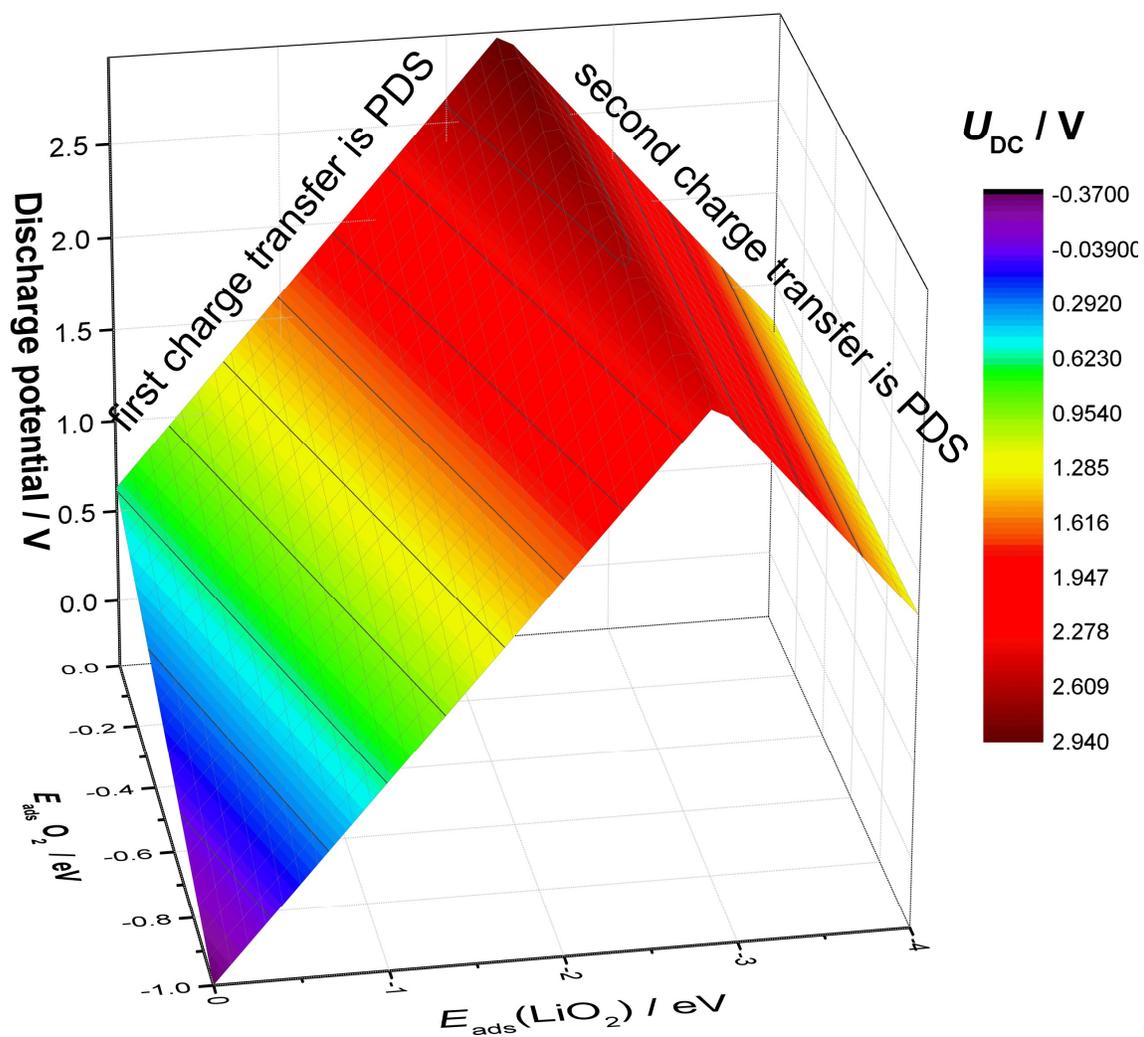

**Figure S6.** 3D view of activity map. Two parts can be identified, depending on the potential determining step for the discharge. For charge process the map would be nearly a reflection of the presented surface with respect to the $U_{DC}$ = 2.96 V plane.